\documentclass[a4paper,12pt]{article}

\usepackage{amsmath,graphicx}
\usepackage{multirow}
\usepackage{amsfonts}
\usepackage{amssymb}
\usepackage{amscd}
\usepackage{cite}
\usepackage{amsmath}
\def\hybrid{\topmargin -20pt    \oddsidemargin 0pt
        \headheight 0pt \headsep 0pt
        \textwidth 6.25in       
        \textheight 9.5in       
        \marginparwidth .875in
        \parskip 5pt plus 1pt   \jot = 1.5ex}

\hybrid

\numberwithin{equation}{section}
\numberwithin{table}{section}

\newcommand{\be}{\begin{equation}}
\newcommand{\ee}{\end{equation}}
\newcommand{\ba}{\begin{aligned}}
\newcommand{\ea}{\end{aligned}}
\newcommand{\M}{{\bf M}}
\newcommand{\bea}{\begin{eqnarray}}
\newcommand{\eea}{\end{eqnarray}}

\def\cF{{\cal F}}
\def\cG{{\cal G}}
\def\Gr{{\rm Gr}}
\def\SO{{\rm SO}}
\def\SU{{\rm SU}}
\def\U{{\rm U}}
\def\nv{n_{\rm v}}
\def\nh{n_{\rm h}}
\def\WW{\omega}

\renewcommand{\Im}{\operatorname{Im}}

\newcommand\rmi{\mathrm{i}}

\def\ax{{\tilde \phi}} 
\def\phet{{\tilde\Phi}} %

\def\ttt{{v}}
\def\t{{t}}

\begin{document}
\begin{titlepage}
\begin{center}
\rightline{\small ZMP-HH/11-22}
\vskip 1cm

{\Large \bf
Heterotic -- Type II duality in the 

hypermultiplet sector}
\vskip 1.2cm
{\bf Jan Louis$^{a,b}$  and Roberto Valandro$^{a}$}

\vskip 0.8cm

$^{a}${\em II. Institut f\"ur Theoretische Physik der Universit\"at Hamburg, Luruper Chaussee 149, 22761 Hamburg, Germany}
\vskip 0.4cm

{}$^{b}${\em Zentrum f\"ur Mathematische Physik,
Universit\"at Hamburg,\\
Bundesstrasse 55, D-20146 Hamburg}
\vskip 0.8cm

{\tt jan.louis,roberto.valandro@desy.de}

\end{center}

\vskip 20mm

\begin{center} {\bf ABSTRACT } \end{center}

\noindent
We revisit the duality between heterotic string theory compactified on
$K3\times T^2$ and type IIA compactified on a Calabi-Yau threefold $X$
in the hypermultiplet sector. We derive an explicit map between 
the field variables of the respective moduli spaces 
at the level of the classical effective actions.
We determine the parametrization of the $K3$ moduli space consistent 
with the Ferrara-Sabharwal form.
From the expression of the holomorphic prepotential we are led to conjecture that both $X$ and its mirror must be $K3$ fibrations in order for the type IIA theory to have an heterotic dual.
We then focus on the region of the moduli space where the metric is
expressed in terms of a prepotential on both sides of the duality.
Applying the duality we derive the heterotic hypermultiplet metric for
a gauge bundle which is reduced to 24 point-like instantons. 
This result is confirmed by using the duality between the heterotic theory on $T^3$ and M-theory on $K3$. We finally study the hyper-K\"ahler metric on the moduli space of an $SU(2)$ bundle on $K3$.

\bigskip

\vfill

\noindent
December 2011

\end{titlepage}
\section{Introduction}

Among the
backgrounds of  different string theories, 
duality relations are conjectured to hold 
\cite{Hull:1994ys,Witten:1995ex}. (For a review see, 
for example, \cite{dewit,Aspinwall:1996mn,Aspinwall:2000fd}.)
One distinguishes perturbative from non-perturbative dualities where
the former hold at weak coupling and can be tested in string 
perturbation theory. T-duality and mirror symmetry are prominent examples
of perturbative dualities.
Non-perturbative dualities  on the other hand involve the string coupling in a
non-trivial way and generically relate weak to strong coupling
physics.
The example we want to focus on in this paper is the heterotic/type
IIA duality, or more specifically the duality between heterotic strings
compactified on $K3\times T^2$ and type IIA theories 
compactified on Calabi-Yau threefolds $X$.
The backgrounds of this duality have 
four infinitely extended Lorentzian ($d=4$) and six compact dimensions.
Eight supercharges are unbroken  or in other words the effective $d=4$ theory 
has $N=2$ supersymmetry.

In $N=2$ supergravity the scalar field space is a direct product of the form
\begin{equation}
  \label{N=2product}
  \M \ =\ {\M}_{\rm  h} \times {\M}_{\rm  v}\ ,
\end{equation} 
where ${\M}_{\rm  h}$ is a $4 n_{\rm  h}$-dimensional quaternionic-K\"ahler
manifold  spanned by the scalars of $n_{\rm  h}$  hypermultiplets,
while ${\M}_{\rm  v}$ is a $2 n_{\rm  v}$-dimensional special-K\"ahler
manifold  spanned by the scalars of $n_{\rm  v}$ vector multiplets \cite{Bagger:1983tt,deWit:1984pk,deWit:1984px,Andrianopoli:1996cm}.
The heterotic dilaton $S$ 
is part of a vector multiplet while the type II dilaton $\phi$
resides in a hypermultiplet. As a consequence  ${\M}_{\rm  v}$ is quantum  
corrected on the heterotic side but `exact' in  type II backgrounds. 
Conversely ${\M}_{\rm  h}$ is exact on the heterotic side 
(if one includes $\alpha'$-corrections) but quantum corrected  
in  type II backgrounds.

The heterotic-type IIA duality has been perturbatively 
well tested for the vector multiplets 
\cite{Kachru:1996pc,Ferrara:1995yx,Klemm:1995tj,Kaplunovsky:1995tm,Antoniadis:1995zn,Aspinwall:1995vk}.
(For a review and further references see, 
for example, \cite{Louis:1996ya,Aspinwall:2000fd}.)
This component is a special K\"ahler manifold described by a holomorphic prepotential ${\cal F}$.  Under the duality the heterotic dilaton $S$ 
is mapped to a geometrical modulus of the Calabi-Yau manifold in the 
type II background.  More precisely, the Calabi-Yau is constrained to be a 
$K3$-fibration where the size of the $\mathbb{P}^1$-base is the type II dual 
of the heterotic dilaton. Thus, weak heterotic couplings corresponds to 
a large  $\mathbb{P}^1$-base -- a limit which can be systematically studied
in type II backgrounds.

On the other hand, much less 
is known for the duality among the hypermultiplets of
${\M}_{\rm  h} $\cite{Aspinwall:1998bw,Aspinwall:1999xs}. 
One of the reasons is that ${\M}_{\rm  h}$ being a
quaternionic-K\"ahler manifold is a more complicated and cannot be
characterized as succinctly as ${\M}_{\rm  v}$  by a holomorphic
prepotential.
However, recently there has been considerable progress in the
understanding of  ${\M}_{\rm  h}$ on the type II side of the duality 
\cite{RoblesLlana:2006ez,RoblesLlana:2006is,Alexandrov:2007ec,RoblesLlana:2007ae,Gaiotto:2008cd,Alexandrov:2008gh,Collinucci:2009nv,Pioline:2009ia,Alexandrov:2011va}
and therefore it seems worthwhile to 
revisit the hypermultiplet sector in the heterotic-type IIA duality \cite{Halmagyi:2007wi}.

The goal of this paper is to assemble further properties of ${\M}_{\rm  h}$ 
on both sides of the duality.
On the type IIA side ${\M}_{\rm  h}$ is spanned by the axio-dilaton, the complex structure moduli of $X$ and the deformations of the RR three-form $C_3$. On the heterotic side, the hypermultiplet scalars are the $K3$ geometric and $B$-field moduli and the deformations of the gauge bundle over $K3$.
First we give the map between the two classical moduli spaces of the dual backgrounds
at the level of the effective action.
To do this, it is convenient to consider an elliptically fibered $K3$ on the heterotic side \cite{Aspinwall:1998bw,Aspinwall:1999xs}.
In order to check the heterotic/type IIA duality
we go to a region of the hypermultiplet moduli space where both heterotic and type IIA theories are in the perturbative regime. This is realized when both $\alpha'$ corrections on the heterotic side and $g_s$ corrections on the type IIA side are negligible. This limit is described in \cite{Aspinwall:1998bw} and we will call it the `double classical limit'. On the heterotic side, it corresponds to taking the elliptically fibred $K3$ with large base and large fibre. 
On the type IIA side, it is realized by taking the dilaton to be large (corresponding to small $g_s$) 
and by making the Calabi-Yau manifold undergo a specific stable degeneration
\cite{Friedman:1997yq,Aspinwall:1998bw}.

At small type IIA string coupling, the quaternionic metric has a special form
in that it is in the image of the c-map \cite{Cecotti:1988qn,FS}.
In this limit the dilaton and universal axion together with 
the deformations of the RR three-form $C_3$ are fibred over a special-K\"ahler base spanned by the complex structure moduli of the Calabi-Yau threefold $X$.
As a consequence the entire quaternionic metric 
is determined by the holomorphic prepotential ${\cal G}$ of the special-K\"ahler base. 
The heterotic hypermultiplet moduli space at leading order in $\alpha'$ has a different structure: First, it is not in the image of the c-map. Second, it has a fibration structure, where the base is spanned by the $K3$ moduli, while the fibre by the bundle moduli. The duality implies that in the double classical limit one recovers both structures on the hypermultiplet moduli space.

Even though the heterotic moduli space is not in the image of the
c-map, this is true for the $K3$ moduli space $SO(4,20)/SO(4)\times
SO(20)$ \cite{Cecotti:1988qn}.
In section~\ref{K3mdsp} we derive the explicit parametrization of this space consistent with the Ferrara-Sabharwal form. This allows us to compare the heterotic metric with the type IIA metric at the level of the prepotential and extract the explicit map of the moduli that gives the match. 
Furthermore, given the form of the prepotential, we are led to conjecture that:
 
\noindent
{\it If type IIA compactified on the CY $X$ has an heterotic dual, then $X$ and its mirror manifold $\tilde{X}$ both have to be a $K3$ fibration.}

With the results just described at hand we are able 
to obtain further information about the heterotic hypermultiplet metric from the type IIA side. The type IIA prepotential depends on all the complex structure moduli of $X$ spanning the special-K\"ahler base. The leading contribution in the degeneration parameter (corresponding to large elliptic fibre of $K3$) is the piece that matches with the $K3$ prepotential. Among the subleading terms we identify the ones related to $\alpha'$ corrections on the heterotic side. Moreover, there are terms in the prepotential that depend on the complex structure moduli corresponding to the heterotic bundle moduli.
We consider in detail the case when the bundle is reduced to 24
point-like instantons 
\cite{Witten:1995gx,Duff:1996rs,Morrison:1996na,Morrison:1996pp,Aspinwall:1997ye,hep-th/9806094}. Using the conjecture on the mirror $\tilde{X}$, we derive the explicit prepotential describing the classical metric on the heterotic side, in the limit of large base and large fibre of the heterotic $K3$. In particular we compute how the $K3$ moduli couple to the bundle moduli.

Finally, we check our results using a different duality which relates
the heterotic string compactified on $T^3$ with  M-theory compactified on $K3$ \cite{Witten:1995ex}. Applying twice the duality map, we are able to confirm the conjecture stated above and to rederive the point-like instanton prepotential. Furthermore, taking the rigid limit of the hypermultiplet moduli space, we find the metric on the moduli space of a smooth $SU(2)$ bundle. To do this we use the fact that in this limit, the wanted prepotential is given by Seiberg-Witten 
prepotential ${\cal F}_{SW}$ \cite{Seiberg:1996nz}.

The paper is organized as follows. Section~\ref{prelim} is
introductory in that
we recall some basic facts about $N=2$ supergravity (section \ref{N=2}),
type II compactified on Calabi-threefolds (section \ref{IICY}),
heterotic compactified on $K3\times T^2$ (section \ref{HetK3})
and the  heterotic-type IIA duality (section \ref{HIIA}).
In section \ref{K3mdsp} we study the $K3$ moduli space and 
give the $K3$ metric in the c-map form. In 
section~\ref{hetIIAdualresults} we focus on the heterotic--type IIA duality.
In 
section~\ref{general} we use the duality with type IIA to derive the structure of heterotic metric, in section~\ref{PLI} we consider the situation of 24 point-like instantons while in section~\ref{3dhypermSec} we studying the heterotic hypermultiplet space using duality with M-theory. 
Finally, in Appendix \ref{FSf} we present the detailed derivation of the result given in section \ref{K3mdsp}, while in \ref{limits} we analyze the two different classical limits.

\section{Preliminaries}\label{prelim}
\subsection{$N=2$ in $d=4$}\label{N=2}
To set the stage 
let us first briefly recall some properties of four-dimensional $N=2$
supergravity (for a review see, for example, \cite{Andrianopoli:1996cm}). The theory consists of a gravitational multiplet, $\nv$ vector multiplets and $\nh$ hypermultiplets. The bosonic components of the gravitational multiplet 
are the spacetime metric $ g_{\mu\nu}, \mu,\nu =0,\ldots,3$
and the graviphoton $A_\mu^0$. A vector multiplet 
contains a vector $A_\mu$
and a complex scalar $t$ as bosonic components. Finally, 
a hypermultiplet 
features four real scalars $q^u$. 
For $\nv$ vector- and $\nh$ hypermultiplets there are thus a total of $2\nv +4\nh$ real scalar fields
in the spectrum. For an ungauged theory the bosonic matter Lagrangian is given by
\begin{equation}\begin{aligned}\label{sigmaint}
{\cal L}\ =\  - \mathrm{i} \mathcal{N}_{IJ}\,F^{I +}_{\mu\nu}F^{\mu\nu\, J+}
+ \mathrm{i} \overline{\mathcal{N}}_{IJ}\,
F^{I-}_{\mu\nu} F^{\mu\nu\, J-}
+ g_{i\bar \jmath}(t,\bar t)\, \partial_\mu t^i \partial^\mu\bar t^{\bar \jmath}
+ h_{uv}(q)\, \partial_\mu q^u \partial^\mu q^v
\ ,
\end{aligned}\end{equation}
where 
$g_{i\bar \jmath},\, i,\bar\jmath = 1,\ldots,\nv$, is 
the metric defined on the $2\nv$-dimensional space ${\M}_{\rm v}$, which
${N}=2$ supersymmetry constrains to be a  special-K\"ahler
manifold  \cite{deWit:1984pk}. This implies
\begin{equation}\label{gdef}
g_{i\bar \jmath} = \partial_i \partial_{\bar \jmath} K^{\rm v}\ ,
\qquad \textrm{for}\qquad
K^{\rm v}= -\ln \rmi\left( \bar X^I \cF_I - X^I\bar \cF_I \right)\ .
\end{equation}
Both $X^I(t)$ and ${\cal F}_I(t)$, $I= 0,1,\ldots,\nv$, are holomorphic functions of the scalars $t^i$ and 
$\cF_I = \partial\cF/\partial{X^I}$
is the derivative of a holomorphic prepotential $\cF(X)$  which is homogeneous of degree two. Furthermore, it is possible to go to a system of `special coordinates' where $X^I= (1,t^i)$.

The $F^{I \pm}_{\mu\nu}$ that appear in
the Lagrangian \eqref{sigmaint} are the self-dual and anti-self-dual parts of
the field strengths of the gauge bosons in
the vector multiplets together with the graviphoton.
Their kinetic matrix $
\mathcal{N}_{IJ}$ is a function of the $t^i $ given by
\begin{equation}
  \label{Ndef}
  {\cal N}_{IJ} = \bar \cF_{IJ} +2\rmi\ \frac{\mbox{Im} \cF_{IK}\mbox{Im}
    \cF_{JL} X^K X^L}{\mbox{Im} \cF_{LK}  X^K X^L} \ ,
\end{equation}
where $\cF_{IJ}=\partial_I \cF_J$. 

The metric on the
$4\nh$-dimensional space  ${\M}_{\rm h}$ is denoted by
$h_{uv}(q),\, u,v=1,\ldots,4\nh,$ in the Lagrangian \eqref{sigmaint}.
It is constrained by 
${N}=2$ supersymmetry to be a quaternionic-K\"ahler manifold  \cite{Bagger:1983tt,deWit:1984px}.
The holonomy group of such manifolds 
is given by $Sp(1)\times Sp(\nh)$
and in addition they admit a
triplet of complex structures $J^x, x=1,2,3$, which satisfy the quaternionic algebra
\begin{equation}\label{jrel}
J^x J^y = -\delta^{xy}{\bf 1} + \epsilon^{xyz} J^z ~.
\end{equation}

A special class of quaternionic-K\"ahler manifolds which will play a
role in the following are the so called `special 
quaternionic-K\"ahler manifolds'.
They can always be constructed from any special K\"ahler manifold $\M_{\rm
  SK}^{2n-2}$ (of dimension $2n-2$) by the c-map \cite{Cecotti:1988qn}
\begin{equation}
  \label{c-map}
c:\quad \frac{SU(1,1)}{U(1)}\ \times\ \M_{\rm SK}^{2n-2} \
\to\ \M_{\rm QK}^{4n}\ .
\end{equation}
An explicit form of the metric on  $\M_{\rm QK}^{4n}$ is known as the
Ferrara-Sabharwal metric which reads \cite{FS}
\begin{equation}\begin{aligned}
  \label{qKmetric}
{\cal L}\  =\ & -(\partial \phi)^2
- e^{4\phi} (\partial \ax +\tilde\xi_A\partial\xi^A - \xi^A\partial\tilde\xi_A )^2
+g_{a\bar b} \partial z^a\partial\bar z^{\bar b}\\
&
+ e^{2\phi} {\Im} {\cal N}^{AB} (\partial\tilde\xi - {\cal N}\partial\xi)_A\overline{(\partial\tilde\xi - {\cal N}\partial\xi)}_B\ ,
\end{aligned}\end{equation}
where the coordinates of the first factor are $(\phi,\ax)$, while
$z^a, a=1,\ldots, n-1$ are
the complex coordinates of  $\M_{\rm SK}^{2n-2}$ with $g_{a\bar b}$ being its
metric. This metric is determined in terms of a holomorphic prepotential $\cG$
by the relation \eqref{gdef} with $\cF$ replaced by $\cG$.
$\xi^A,\tilde\xi^A, A=0,\ldots,2n-2$ are the $2n$ real coordinates of a
complex torus which is fibred over  $\M_{\rm SK}^{2n-2}$.
Their couplings ${\cal N}_{AB}$ are determined in terms of the holomorphic
prepotential $\cG$ of $\M_{\rm SK}^{2n-2}$ via
\begin{equation}
  \label{Mdef}
{\cal N}_{AB} = \bar \cG_{AB} + 2\rmi\ \frac{{\Im}(\cG_{AC}) Z^C{\rm Im}(\cG_{BD}) Z^D}{{\Im}(\cG_{CD}) Z^CZ^D}\ ,
\end{equation}
where $Z^A = (1,z^a)$ are the homogeneous coordinates on $\M_{\rm SK}^{2n-2}$.

\subsection{Type II compactified on Calabi-threefolds}\label{IICY}

Calabi-Yau manifolds $X$ have  $h^{1,1}(X)$ K\"ahler moduli and  $h^{1,2}(X)$
complex structure moduli which can be varied independently. Hence the
total moduli space is a product
\begin{equation}
  \label{CYmoduli}
  \M(X) \ =\ {\M}_{(1,1)}(X) \times {\M}_{(1,2)}(X)\ ,
\end{equation}
with each component being a special K\"ahler manifold.
In type IIA one has ${\M}_{(1,1)}={\M}_{\rm  v}$ which,
as discussed in the previous section, is characterized by a holomorphic ${\cal F}_{\rm II}$ of the generic form
\begin{equation}\label{Fvectorgen}
\cF_{\rm II} =  \rmi d_{ijk} t^i t^j t^k+ c + {\cal O} (e^{- 2\pi \rmi t^j})
\ .
\end{equation}
The $t^i$ now denote the K\"ahler moduli of $X$ with $d_{ijk}$ being
the classical intersection numbers of their dual two-cycles. $c$ is a
constant proportional 
to the Euler number of $X$ while the last term denotes the worldsheet instanton corrections.

On the other hand 
${\M}_{(1,2)}\subset{\M}_{\rm  h}$ where the full ${\M}_{\rm  h}$ also
features the dilaton and axion plus $2(h^{1,2}+1)$ scalars
from the RR-sector.
 At the string tree level 
${\M}_{\rm  h}$ is a special quaternionic-K\"ahler manifold in the image of the c-map \eqref{c-map}, i.e.\ the metric is of the form \eqref{qKmetric}.
Now $(\phi,\ax)$ are the dilaton and  axion respectively while the 
$z^a, a=1,\ldots,h^{1,2}$,
are the complex structure moduli parametrizing  the special K\"ahler
manifold ${\M}_{(1,2)}$ with metric $g_{a\bar b}$. 
In full generality this metric is only known for Calabi-Yaus with 
$h^{1,2}$ small.
The real
$(\xi^A,\tilde\xi^A), A=0,\ldots,h^{1,2},$ arise from expanding the
RR three-form $C_3$
and are the fibre coordinates
(coordinates of the intermediate Jacobian). So  altogether there are indeed
$4(h^{1,2}+1)$ scalars $(\phi,\ax,z^a,\xi^A,\tilde\xi^A)$.

In type IIB the assignment is reversed with 
${\M}_{(1,2)}={\M}_{\rm  v}$ while ${\M}_{(1,1)}\subset{\M}_{\rm  h}$.
The full ${\M}_{\rm  h}$ is again completed by the 
axio-dilaton together with $2(h^{1,1}+1)$ scalars
from the RR-sector.
Mirror symmetry states that 
 type IIA compactified on $X$ is equivalent 
to type IIB compactified on the mirror manifold $\tilde X$.
This implies in particular
\be
{\M}_{(1,1)}(X) = {\M}_{(1,2)}(\tilde X)\ ,\qquad 
{\M}_{(1,2)}(X) = {\M}_{(1,1)}(\tilde X)\ .
\ee
or equivalently
\be\label{FGmirror}
\cF_{\rm IIA}(X) = \cG_{\rm IIB}(\tilde X)\ ,\qquad 
\cG_{\rm IIA}(X) = \cF_{\rm IIB}(\tilde X)\ .
\ee

Since the dilaton is in the hypermultiplet sector the metric \eqref{qKmetric}
receives quantum correction which generically cannot be written in the
form \eqref{qKmetric}. In other words the quantum corrected hypermultiplet 
sector  is  no longer a special
quaternionic-K\"ahler manifolds and the quantum corrections cannot be viewed
as correction of the holomorphic prepotential $\cG_{\rm II}$ \cite{Strominger,Gunther:1998sc,Antoniadis:1997eg,AMTV,RoblesLlana:2006ez}.

\subsection{Heterotic compactified on $K3\times T^2$ }\label{HetK3}

On the heterotic side the vector multiplets arise from the original
ten-dimensional gauge group $E_8\times E_8$ or $SO(32)$ together with 
three Kaluza-Klein vector fields of $T^2$.\footnote{The fourth vector field is the graviphoton.}
Their couplings are described by a prepotential of the form \cite{deWit:1995zg,Antoniadis:1995ct,Louis:1996mt}
\begin{equation}\label{Hvector}
 \cF_{\rm het}= \rmi S\eta_{\alpha\beta}T^\alpha T^\beta 
+ f(T) +g(T,W)+ {\cal O} (e^{- 2\pi \rmi S}) 
\ , 
\end{equation}
where $S$ denotes the heterotic dilaton and $\eta_{\alpha\beta}$ is an $SO(2,n-2)$ metric ($\alpha,\beta=1,...,n$). The $T^\alpha$ 
consist of two moduli parametrizing K\"ahler and complex structure
deformations of the $T^2$ together with Wilson-line moduli
in the Cartan subalgebra
of the  $E_8\times E_8$ or $SO(32)$  which parametrize the Coulomb branch of
this gauge bundle.
 $f(T)$ are the one-loop threshold corrections 
(computed for example in \cite{deWit:1995zg,Antoniadis:1995ct,Harvey:1995fq}) while 
$g(T,W)$ denotes the couplings of additional gauge multiplets $W$
which are light due to non-perturbative effects~\cite{Louis:1996mt}. 
Finally, ${\cal O} (e^{- 2\pi \rmi S})$ stands for non-perturbative corrections which are governed by the dilaton.
The string tree level corresponds to large $S$ 
in which case only the first term in \eqref{Hvector} survives and 
the vector multiplets parametrize the space 
\begin{equation}
  \label{Vclassical}
 \M_{\rm v}= \frac{\SU(1,1)}{\U(1)}\times\ 
\frac{\SO(2,n-2)}{\SO(n-2)\times \SO(2)}\ .
\end{equation}
Among the one-loop and non-perturbative corrections one finds
the Seiberg-Witten ${\cal F}_{\rm SW}$ in an appropriate double scaling 
limit where 
all gravitational corrections are turned off \cite{hep-th/9508155,Lerche:1996xu}.
(We return to this point section~\ref{HIIA}.)

The hypermultiplet sector 
sensitively
depends on choice of gauge bundle and is 
constrained by the Bianchi identity
\be\label{Bianchi}
\int_{\rm K3} {\rm tr} F^2  + n_5 = n+n_5=\int_{\rm K3} {\rm tr} R^2 = 24\ ,
\ee
where $n_5$ counts the number of background five-branes and 
$n=\int_{\rm K3} {\rm tr} F^2 $ is the instanton number.
In perturbative compactifications  $n_5=0$ holds.
The 
hypermultiplet scalars are 
the $K3$~moduli (or rather a subset thereof), the moduli arsing from
the NS $B$-field and
the moduli of instanton solutions on $K3$, i.e. 
the gauge bundle deformations. In fact the space of gauge bundle
moduli is fibred over the moduli space of a $K3$
base.\footnote{Strictly speaking there also is the possibility of
  charged matter multiplets. However, for most of our considerations
  we go to the Higgs- or Coulomb branch of these theories where they
  become massive or can be viewed as additional gauge neutral bundle
  moduli. 
Some couplings of the charged matter fields have been determined
recently in ref.~\cite{schasny}.}

At specific loci of its moduli space the $K3$ becomes singular.
These singularties follow an ADE-classification.
Near such a singularity one can replace the $K3$ by an ALE space
which is asymptotic to $\mathbb{R}^4/\Gamma$ where the finite group $\Gamma$ depends on the choice of the ADE-singularity. 
In \cite{Witten:1999fq,Rozali:1999va,Aspinwall:1999xs,Mayr:1999bk}
it was argued that in this limit the hypermultiplet moduli space 
coincides with the moduli space of a three-dimensional gauge theory
with the corresponding ADE-gauge group.

On the heterotic side the dilaton is part of a vector multiplet
and hence ${\M_{\rm h}}$ receives $\alpha'$-corrections but 
is ``exact'' in string ($g_s$) perturbation theory.
However, this argument is too naive since non-perturbative effects
which do not have the canonical $e^{-g_s^{-2}}$ dependence can contribute
to ${\M_{\rm h}}$. 
In fact the moduli space of instanton solutions is generically singular when
the size $\rho$ of an instanton shrinks to zero, i.e.\ 
$\rho\to 0$ \cite{Callan:1991dj}. This singularity can be seen in string theory
already at the tree level and in particular in the heterotic
hypermultiplet moduli space.
In ref.~\cite{Witten:1995gx,Duff:1996rs} the singularity was interpreted as  a
non-perturbative effect -- either as an enhanced gauge symmetry for 
the $SO(32)$ heterotic string or as a wrapped 5-brane becoming
 tensionless for the $E_8\times E_8$ heterotic string.

For an arbitrary gauge group $G$ the moduli space of instantons on $K3$
is a hyper-K\"ahler manifold with (quaternionic) dimension
\be
{\cal N}_n = n h(G)-{\rm dim}(G)\ ,
\ee
where $n$ is the instanton number and $h(G)$ the dual Coxeter number of $G$.
For a small $E_8$ instanton $n$ changes by one
and thus $ h(E_8)-1=29$ hypermultiplets are fixed where the extra modulus parametrizes
the location of the small instanton.
In $d=6$  this modulus is part of an additional tensor multiplet 
which
in $d=4$ can be dualized to an additional $U(1)$ vector
multiplet.\footnote{This multiplet is know as the $N=2$ 
vector-tensor multiplet 
\cite{deWit:1995zg} which is dual to vector multiplet.}
These $U(1)$ vector multiplets are denoted by $W$ in \eqref{Hvector} 
and some of their couplings have been computed, for example in ref.~\cite{Louis:1996mt}.

For a generic instanton background with structure group $SU(n)$, we
can describe its moduli by means of the spectral cover
\cite{Friedman:1997yq}. In fact, $SU(n)$ bundles over an elliptically
fibred manifold 
are described by a spectral cover $\mathcal{C}_H$ (i.e. an $n$-cover of the base 
$\pi: \mathcal{C}_H \rightarrow B$) and a line bundle $\mathcal{N}$ over $\mathcal{C}_H$
\cite{Friedman:1997yq}.  The line bundle over $\mathcal{C}_H$ is given in terms of its first Chern class $c_1(\mathcal{N})$ and a 
twist by a flat bundle on $\mathcal{C}_H$ (if it is not simply-connected).
When the elliptic fibration is a $K3$ surface, the spectral cover is a curve that in general has non-trivial fundamental group and the first Chern class 
of $\mathcal{N}$ is completely determined by the spectral curve. In fact \cite{Friedman:1997yq},
$$
  \mathcal{N}=K_{\mathcal{C}_H}^{1/2}\otimes K_B^{-1/2}\otimes
  \mathcal{F}\ ,
$$
where $K_{\mathcal{C}_H}$ and $K_B$ are the canonical bundles of $\mathcal{C}_H$ and $B$ and $\mathcal{F}$ is a flat bundle over $\mathcal{C}_H$.
Such flat bundles are classified by the Jacobian $J(\mathcal{C}_H)$ of $\mathcal{C}_H$.

\subsection{ Heterotic-type IIA duality}\label{HIIA}

In \cite{Hull:1994ys,Witten:1995ex} it was conjectured that apart from
perturbative dualities also non-perturbative dualities
hold among string backgrounds. In these cases the dualities involve the
string coupling in a non-trivial way.
One of the prominent example is the duality of the heterotic string
compactified on a four-torus $T^4$ and type IIA string theory
compactified on $K3$.
These backgrounds have six Lorentzian space-time dimension and 16
unbroken supercharges. The respective dilatons are mapped inversely 
to each other mapping the strong coupling region of one theory to
the weak coupling regime of the dual theory.

A close ``cousin'' of this duality is conjectured to hold 
in four space-time dimensions
with eight unbroken supercharges and can be viewed as a fibred version
of the six-dimensional duality.
More precisely, one fibres both sides over a 
$\mathbb{P}^1$ base such that on the heterotic side the
compactification manifold is $K3\times T^2$ with the $K3$ 
being elliptically fibred 
(with a section) over the $\mathbb{P}^1$. (We denoted the
$\mathbb{P}^1$ on the heterotic side by $B_H$ and the $K3$ 
by $S_H$ in the following.) 
On the type II side the $K3$ is fibred 
over the $\mathbb{P}^1$ base (which we denote by $B_{\rm II}$ in the following)
in such a way that the resulting threefold $X$ is Calabi-Yau. 
Due to the elliptic fibration of  the $K3$ the threefold $X$  is also
elliptically fibred over a ruled surface $\Sigma$.
Thus the conjectured duality relates
the heterotic string
compactified on $K3\times T^2$ with an elliptic $K3$ to  type IIA string theory
compactified on a $K3$-fibred Calabi-Yau threefolds
\cite{Kachru:1996pc,Ferrara:1995yx,Klemm:1995tj,Aspinwall:1995vk}. 
The heterotic dilaton, which is part of a vector multiplet, 
is mapped to the type IIA K\"ahler modulus which controls the size 
of the $B_{\rm II}$ and thus it is not a
strong-weak duality but rather relates the string coupling to a
geometrical modulus. This picture has been confirmed in the vector
multiplet sector by organizing the type II vector
multiplet couplings in a form corresponding to the heterotic
perturbation theory and then comparing both sides for a (large) number of dual
backgrounds.\footnote{See, for example,
  \cite{Kachru:1996pc,Ferrara:1995yx,Klemm:1995tj,Aspinwall:1995vk,Kaplunovsky:1995tm,Antoniadis:1995zn} or the reviews  \cite{Louis:1996ya,Klemm:2005tw}.}
The success of these checks is partly related to the fact that the $N=2$
vector multiplet couplings are encoded in terms of the
holomorphic prepotential ${\cal F}$ 
which can be easily compared or constrained.

Concretely one finds in the large dilaton/large base volume limit
\begin{equation}\label{Fvector}
\cF_{\rm II} = \cF_{\rm het}=  \rmi S\eta_{\alpha\beta}T^\alpha T^\beta 
+ \rmi d_{\alpha\beta\gamma} T^\alpha T^\beta T^\gamma+\ldots
\ ,
\end{equation}
where for simplicity we have adopted the heterotic notation. On the type IIA side,
$S$ denotes the volume modulus
of the base $B_{\rm II}$, while
the $T^\alpha$ denote the K\"ahler moduli of the
generic $K3$~fibre. $\eta_{\alpha\beta}$ is an $SO(2,n-2)$ metric while
$d_{\alpha\beta\gamma}$
are the (classical) intersection numbers of the K\"ahler moduli on the
type II side which are related to one-loop threshold corrections on the
heterotic side.
The ellipsis stand for exponentially suppressed terms and moduli
parametrizing singular $K3$ fibres (see \cite{Klemm:1995tj,Aspinwall:1995vk,Kaplunovsky:1995tm} for  more details).

At special points in this moduli space  non-Abelian gauge bosons 
occur. On the heterotic side this is a standard Higgs mechanism where
all (or part) of the original $E_8\times E_8$ gauge symmetry becomes visible.
On the type II side
the non-Abelian gauge enhancement occurs in the vicinity of 
the ADE-singu\-larities of the $K3$ fibres.
In an appropriate ``double scaling limit'' which corresponds to turning off gravity and zooming in on one specific singularity  one can replace the $K3$ fibre
by an ALE space and recover the Seiberg-Witten $\cF_{\rm SW}$ from 
$\cF_{\rm II}$ \cite{hep-th/9508155,Lerche:1996xu}.

A similar analysis for the hypermultiplet sector was pioneered in 
 \cite{Aspinwall:1998bw} where on the heterotic side 
the dual type II dilaton was identified. However, due to the more
complicated
geometry of the hypermultiplets the duality has not been established
at a similar quantitative level and one of the goals of this paper is
to improve this situation.  In order to do so we need
however to first recall the results of \cite{Aspinwall:1998bw}.
For simplicity we confine our attention to $E_8\times E_8$ backgrounds henceforth.

As a first step to repeat the analysis 
in the hypermultiplet sector ref.~\cite{Aspinwall:1998bw} 
identified the volume of the heterotic base $B_H$ 
with the type IIA dilaton. 
With this identification it is in principle possible to 
formally organize the heterotic hypermultiplet couplings in a type
IIA perturbation theory, or in other words in powers of 
the type IIA dilaton. The problem is that on the heterotic side 
very little is explicitly known about $\M_{\rm h}$.

In the duality, type IIA tree-level contributions correspond to $B_H$ being large.
In this  limit one can still have 
$\alpha'$ corrections on the heterotic side, as they depend
generically on the volume of $K3$.
Therefore one can achieve a further simplification by considering 
 $K3$s with large volume and large $B_H$. 
A sufficient condition for a generic $K3$ (i.e.\ without ADE singularities) with large base to have
large volume is to also take the elliptic fibre to
be large. 
The modulus parametrizing the (complexified) fibre volume
corresponds to a specific 
complex structure modulus of the Calabi-Yau threefold $X$ 
that we will call the ``smoothing
parameter'' $\sigma$. 
Taking the volume of the fibre to be infinitely large then
corresponds to a particular limit (let us call it $\sigma\rightarrow \infty$)
in the complex structure moduli space of $X$ which is known as  
the stable degeneration limit $X\to X^\sharp$. In this limit
$X^\sharp$ is still a $K3$ fibration over $B_{\rm II}$, but now the $K3$ is degenerate over each point: It is the union of two
rational elliptic surfaces intersecting along an elliptic curve.
 Correspondingly, $X^\sharp$ is the union
of two three-folds $X_1$ and $X_2$, intersecting along an elliptically fibred $K3$ surface $S_\ast$.
The type IIA complex structure moduli describing the 18 complex
deformations of $S_\ast$ are mapped by duality to the complex structure moduli of
the heterotic $S_H$. On the other side, the type IIA moduli describing the complex structure deformation of $X_1$ and $X_2$ are mapped to the complex deformations of the two spectral curves describing the bundle moduli of the two $E_8$ factors.

Let us now consider the third cohomology group of $X^\sharp$. It splits into three pieces \cite{Aspinwall:1998bw}:
\begin{equation}
  \label{H3}
 H_3(X^\sharp) \to H_3(X_1) +  H_3(X_2)+ M_{20}\ .
\end{equation}
The first and the second subspaces are the spaces of three-cycles of $X_1$ and of $X_2$ respectively; they are disjoint as
there are no three-cycles in $S_\ast$. The third subspace in \eqref{H3} is the set of three-cycles constructed by two-cycles 
of $S_\ast$: Since for generic complex structure the only $(1,1)$-cycles of $S_\ast$ are the fibre and the base
of the elliptic fibration, there are 20 two-cycles of $S_\ast$ that are homologically trivial in  $X^\sharp$ and that are
the boundary of a three-chain in $X_1$ and a three-chain in $X_2$. By these chains one can construct closed
three-cycles in $X^\sharp$.

Going away from the degenerate limit, a set of three-cycles blow up 
which we will call $M_{20}^\star$. In fact, as $\sigma$
becomes finite, a circle fibred over $S_\ast$ appears (it was shrunk to zero in the degenerate limit). The fibration
of this circle over the 20 two-cycles orthogonal to base and fibre of $S_\ast$ gives
the three-cycles in $M_{20}^\star$. Together with the three-cycles of
$M_{20}$, they form a symplectic subspace of $H_3(X^\sharp)$.

On the type IIA side 
$H_3(X)$ determines the complex structure deformations $z^a$
but also the $2(h^{1,2}+2)$ real
moduli $\xi^A,\tilde\xi^A, A=0,\ldots,h^{1,2}$
which arise as zero modes of the RR three-form
$C_3$. 
In the dual heterotic vacuum the  $C_3$-moduli arising from 
$H_3(X_1)$ correspond to
bundle moduli of the first $E_8$, while the $C_3$-moduli arising from $H_3(X_2)$ correspond to
bundle moduli of the second $E_8$. In particular, they correspond to the Jacobians of the two spectral curves related to the two $E_8$ factors. Furthermore, the $C_3$-moduli
arising from $M_{20}$ are mapped to the heterotic $B$-field along two-cycles of $S_H$. Away from the
degenerate limit there are also the moduli given by $C_3$ along
$M_{20}^\star$.
These correspond to
the K\"ahler moduli (with exclusion of fibre and base volumes) of the heterotic $S_H$. They are frozen in the limit of large fibre and base in heterotic theory.

Let us summarize the map of the spectrum in table \ref{tab:map}. In
particular,
note that we can give a type IIA counterpart for all heterotic $K3$
moduli.

\bigskip

\begin{table}[ht]\begin{center}
\begin{tabular}{|c |c|c |c|c|}
\hline
 \multicolumn{2}{|c|}{IIA} &
   \multicolumn{2}{|c|}{Het} & $\#_{\mathbb{R}}$\\
\hline\hline
dilaton/axion& $\phi,\ax$& 
vol($\mathbb{P}^1$) + $B$-field & $\phet$ &2\\
\hline
smoothing parameter& $\sigma$ & vol(fibre) + $B$-field &$s$&2\\
\hline
$cs(S_*)$ &$\tau$ & cs($S_H$) & $t$&36\\
\hline
$C_3(M_{20})$ & $\tilde{\xi}$& 20 $B$-fields&$\tilde c$&20\\
\hline
$C_3(M^\star_{20})$ & $\xi$& 20 $J$ moduli  &$c$&20\\
\hline
$cs(X_1) $&& $E_8^1$-bdl moduli: spectral cover $\mathcal{C}^1_H$ &&model dep.\\
\hline
$C_3(X_1)$&& $E_8^1$-bdl moduli: Jacobian $J(\mathcal{C}^1_H)$ &&model dep.\\
\hline
$cs(X_2) $&& $E_8^2$-bdl moduli: spectral cover $\mathcal{C}^2_H$ &&model dep.\\
\hline
$C_3(X_2)$&& $E_8^2$-bdl moduli: Jacobian $J(\mathcal{C}^2_H)$ &&model dep.\\
\hline
\end{tabular}
\caption{Heterotic-type II map at the tree-level.}\label{tab:map}
\end{center}\end{table}

\section{$K3$ moduli space in Ferrara-Sabharwal form}\label{K3mdsp}

One of our goals is to make (part of) the table~\ref{tab:map} explicit
at the level of the effective action. 
In order to do so we start at 
the type IIA tree level and take the stable degeneration
limit which allows us to neglect $\alpha'$ corrections in 
the dual heterotic background.
Furthermore, in this section we freeze all  gauge bundle moduli for simplicity
and only study the `geometrical' moduli space  arising from deformations of
the $K3$ metric together with the moduli arising from the NS
$B$-field. For this subset of fields 
we determine the precise map between
heterotic and type II field variables.

It is convenient to introduce the notation
\begin{equation}
  \label{Gr}
 \Gr_{n,m}\ \equiv\ \frac{\SO(n,m)}{\SO(n)\times \SO(m)}\ ,
\end{equation}
which has real dimension  $dim_{\bf R}(\Gr_{n,m})= nm$.
The 58 moduli of the $K3$ metric form the space
\be
 \Gr_{3,19}\times R^+\ ,
\ee
where the $R^+$ corresponds to the volume.
Including the 22 $B$-fields
the moduli space is 
locally given by \cite{Seiberg:1988pf}
\begin{equation}
  \label{K3mod}
  \M_{K3}\ =\ \Gr_{4,20}  
\ .
\end{equation}

However, it can happen that a specific
solution of \eqref{Bianchi} fixes some of the $K3$ moduli.
Since $N=2$ constrains $\M_h$ to be quaternionic-K\"ahler one  typically
obtains the  $4n$-dimensional subspace 
\begin{equation}
  \label{K3modcheck}
  \check\M_{K3}\ =\ \Gr_{4,n} 
\subset \M_{K3}
\ ,
\end{equation}
where $n\le 20$.\footnote{For the standard embedding 
and the case of point-like instantons
\cite{Witten:1995gx,Aspinwall:1997ye}, 
all $K3$ moduli can be varied freely and one has $n=20$.
In the following we always discuss the case of arbitrary $n$ however.}
If we additionally freeze the $B$-fields $\check\M_{K3}$ is reduced to 
$\Gr_{3,n-1}\times R^+$. Fixing also the K\"ahler class of $K3$
a further reduction to $\Gr_{2,n-1}$ occurs. Finally, if the $K3$ is also
elliptically fibred the complex structure 
moduli preserving the elliptic fibration span 
$\Gr_{2,n-2}$. This can be summarized
by the chain
\begin{equation}
  \label{Grcs}
 \check\M_{K3}\ =\ \Gr_{4,n}\ \to\ 
\Gr_{3,n-1}\times R^+\ \to\ \Gr_{2,n-1}\  \to\   \Gr_{2,n-2}\ .
\end{equation}

The heterotic action for the $K3$ moduli can be written as 
\cite{Duff:1995wd,Haack:2001iz}
\begin{equation}\begin{aligned}
  \label{Lhet}
{\cal L}\  =\ & -\tfrac18(\partial \rho)^2 + \tfrac1{16}
(\partial \hat{M})^2 -\tfrac14 e^{\rho} \hat{M}^{IJ}\partial B_I\partial B_J\ ,
\end{aligned}\end{equation}
where $e^{-\rho}$ is the $K3$ volume, 
$\hat M$ is a $\SO(3,n-1)$ matrix parametrizing  $\Gr_{3,n-1}$
 and $B_I$ are  $n+2$ $B$-fields.
In order to prepare the comparison with type IIA we need to 
rewrite this action in the Ferrara-Sabharwal form \eqref{qKmetric}.
Indeed it is known \cite{Cecotti:1988qn} that $\Gr_{4,n}$ is a special
quaternionic manifold, i.e.\ 
this space is in the image of the c-map with a special
K\"ahler base $\M_{\rm SK}^{2(n-1)}= \frac{SU(1,1)}{U(1)} \times
\Gr_{2,n-2}$.
So we only need to identify the base inside  $\Gr_{4,n}$
and then put \eqref{Lhet} into the form \eqref{qKmetric}.
However, this turns out to be surprisingly involved. 

As we reviewed in section~\ref{N=2} 
the Ferrara-Sabharwal metric   \eqref{qKmetric} in general 
is a metric  of a special quaternionic-K\"ahler manifold $\M_{\rm
  QK}^{4n}$
which is in the image of the c-map \eqref{c-map}. In other words 
it has a base $\frac{SU(1,1)}{U(1)}\ \times\ \M_{\rm SK}^{2n-2}$ where
the first factor is 
parametrized by the coordinates $(\phi,\ax)$  (corresponding to
dilaton and axion on the type II side),  while the second factor 
is parametrized by the complex variables $z^a$.
Our first task therefore is to identify this base in the heterotic
metric \eqref{Lhet} in the stable degeneration limit.

In order to do so we 
only keep the two K\"ahler moduli $\phet, s$ corresponding to
the volume of the base $B_H$
and the elliptic fibre 
(first two lines in table \ref{tab:map}) and for the moment 
freeze the other $n$ K\"ahler moduli.  As noted
in \eqref{Grcs} the  $2(n-2)$ complex structure deformations $\t^i$
which preserve this elliptic
fibration  span the moduli space $\Gr_{2,n-2}$. (They correspond to
the third line in table \ref{tab:map}.) 
Thus  altogether we have in this limit
\begin{equation}
  \label{Grhere}
 \check\M_{K3}\ \to\ \frac{SU(1,1)}{U(1)} \times \M_{\rm SK}^{2(n-1)} \ ,\qquad \textrm{with}
\qquad
\M_{\rm SK}^{2(n-1)}\ =\ \frac{SU(1,1)}{U(1)} \times \Gr_{2,n-2}\ ,
\end{equation}
where the two $\frac{SU(1,1)}{U(1)}$--factors are spanned by $\phet$ and $s$.
From \eqref{c-map} we now see that we can
reconstruct the entire  $\check\M_{K3} = \Gr_{4,n}$ as a 
c-map of the special-K\"ahler base $\M_{\rm SK}^{2(n-1)}$. 
The $2n$-dimensional space which  is fibred over $\M_{\rm SK}^{2(n-1)}$ is
parametrized by the $n$ K\"ahler moduli which were frozen so far
and $n$ $B$-fields.

In appendix~\ref{FSf} we perform a number of field redefinitions
which at the end put \eqref{Lhet} into the form
\eqref{qKmetric}. 
A complication arises from the fact that the metric of 
$\M_{\rm SK}^{2(n-1)}$ 
(denoted by $g_{ab}$ in \eqref{qKmetric}) is not immediately
in the canonical form
as given in \eqref{gdef}. Rather it is in field variables where the 
holomorphic prepotential does not exist and only after a symplectic
rotation can be expressed as the derivative of a prepotential ${\cal
  G}$
\cite{Ceresole:1995jg}.
After this rotation we find (as expected)
\begin{equation}\label{prepMir}
 \cG_{\rm het} = \rmi s\,\def\ttt{{v}}
 \eta_{ij}t^it^j \ , \quad i,j=2,\ldots,n-1\ ,
\end{equation}
where
the $\t^i$ are the $(n-2)$ complex structure deformations preserving the
elliptic $K3$ and $\eta_{ij}$ is an $\SO(2,n-2)$ metric.

\section{Heterotic - Type IIA duality in the hypermultiplet sector}\label{hetIIAdualresults}

\subsection{General considerations}\label{general}

Let us now see what in turn the heterotic results imply on the type
IIA side. From \eqref{prepMir} we conclude 
that in the stable degeneration limit the complex structure
moduli space of $X$ has to take a particular form. There has to be a
limit where the holomorphic prepotential $\cG_{\rm II}$ which appears in the Ferrara-Sabharwal metric and which describes the complex structure moduli space
of $X$ takes a
form identical to \eqref{prepMir}, i.e.\
\begin{equation}\label{prepMircc}
 \cG_{\rm II} = \rmi \sigma \eta_{ij}\tau^i\tau^j + \ldots\ ,\qquad i,j=2,\ldots,n-1\ ,
\end{equation}
where $\sigma,\tau^i$ are now complex structure moduli of the dual type IIA background.  However, the form \eqref{prepMircc} generically does not appear in the 
complex structure moduli space of $X$.
In fact comparing \eqref{prepMircc} with \eqref{Fvector} we see that $\cG_{\rm II}$ coincides
with $\cF_{\rm II}$ in the large volume (or more precisely large
$B_{\rm II}$ limit)
if we identify $\sigma \leftrightarrow S, \tau^i \leftrightarrow T^\alpha$.
This implies that in the complex structure moduli space there is a
limit where the prepotential  $\cG_{\rm II}(X)$ agrees with
$\cF_{\rm II}(\tilde X)$ in the K\"ahler moduli space 
of some other (mirror) manifold $\tilde X$. 
Of course this is precisely the
statement of mirror symmetry and
the necessary agreement  of \eqref{Fvector} with \eqref{prepMircc}
suggests that the 
mirror  threefold $\tilde X$ also has to be a $K3$~fibration.
We are thus lead to the conjecture:

{\sl Type II compactified on a Calabi-Yau threefold
$X$ can be dual to a heterotic string vacuum only if $X$ and its 
mirror threefold $\tilde X$ are simultaneously $K3$~fibrations.}

The modulus $\sigma$ is than the mirror dual of the volume $S$ of the base 
$B_{\rm II}$ of $\tilde X$ and the $\tau^i$ are the mirror duals
of the $K3$~fibre K\"ahler moduli $T^\alpha$ of  $\tilde
X$.\footnote{Note that the set of Calabi-Yau three-folds  studied in
  \cite{Aspinwall:1999xs} do satisfy this conjecture.}

Once we assume the validity of this conjecture we can draw further conclusions
in that the structure of the vector multiplet couplings of the $K3$-fibred $\tilde X$ has additional properties.  For example,
the Coulomb branch has a  non-Abelian gauge enhancement
 at the ADE-singularites of $K3$ as we briefly discussed in section~\ref{HetK3}.
Thus in an appropriate  double scaling limit one has 
$\cG_{\rm II} = \cG_{\rm SW}(ADE)$ \cite{Witten:1999fq,Rozali:1999va,Aspinwall:1999xs,Mayr:1999bk}.

At a finite number of points on $B_{\rm II}$
the $K3$ fibre can degenerate and the K\"ahler moduli associated with
the components of such degenerate fibres  have no intersection
with the $B_{\rm II}$ volume~$S$~\cite{Aspinwall:1995vk}. Translated to the complex structure
moduli space of $X$ this says that  $\cG_{\rm II}$
can be parametrized analogous to \eqref{Hvector} by
\begin{equation}\label{prepMirc}
 \cG_{\rm II} = \rmi \sigma \eta_{ij}\tau^i\tau^j + f(\tau) +g(\tau,\WW)+\ldots\ , 
\end{equation}
where the ellipsis stand for terms which are exponentially suppressed
in the large $\sigma$ limit. $f(\tau)$ is a function of the $\tau^i$
which in the large complex
structure limit is cubic plus a constant $\varrho$ proportional to the Euler number
plus exponentially suppressed terms
\be\label{foftau}
f(\tau) = \rmi d_{ijk} \tau^i\tau^j \tau^k + \varrho+ {\cal O} (e^{- 2\pi \rmi \tau^i})\ .
\ee

This in turn further determines the structure of $\cG_{\rm het}$ of which we only 
computed the leading contribution in the large $s$-limit in 
\eqref{prepMir}. The couplings of $f(\tau)$ given in \eqref{foftau}
we can understand on the heterotic side as $\alpha'$ corrections. Recall
that the heterotic $t^i$ (which are dual to the $\tau^i$) correspond to
complex structure deformations of $K3$ that preserve the elliptic fibration.
Furthermore, they control the volume of the transcendental two-cycles $C_2$
which is approximately vol$(C_2)\propto e^{-\rho/2}\,t$. In the limit of large 
$K3$ volume (i.e.~$e^{-\rho/2}$ large)
and for generic values of $t^i$, these cycles are also large. 
However, when some $t^i$ go to zero, a two-cycle shrinks 
and a worldsheet instanton wrapping $C_2$ contributes to the prepotential. 
Thus we can identify the last term in \eqref{foftau} as a contribution from worldsheet instantons on the heterotic side.\footnote{Note that this term appears already at the classical level on the type IIA side.}
The cubic term  in \eqref{foftau} is a bit more difficult to understand since
it appears as a perturbative $\alpha'$ correction. 
On the type II side it becomes relevant when $\sigma$ is finite, i.e.\ 
when we go away from the degeneration limit in type IIA. Departing from 
this limit on the heterotic side means that the $\alpha'$ 
corrections become relevant. 
However, the cubic term cannot be interpreted as an instanton correction, as it becomes subleading when the volumes of the wrapped cycles become small.
It would be nice to identify the origin of this term in more detail on the heterotic side.

The function
$g(\tau,\WW)$ in \eqref{prepMirc} 
summarizes on the type II side 
the contribution of the degenerate fibres whose moduli 
we denote by $\WW$. As
already stated above it cannot have any $\sigma$-dependence.
Furthermore, they are related to vector multiplet couplings 
of the type IIA theory on the mirror $\tilde X$.
Thus using the results of \cite{Louis:1996mt}
we can give generically 
\be\label{gtau}
g(\tau,\WW)=c_{ij}\tau^i\tau^j \WW + c_i\tau^i \WW^2 + c \WW^3 + \ldots\ ,
\ee
where $c_{ij}$, $c_i$ and $c$ are model dependent constants.

\subsection{Point-like instantons}\label{PLI}

We now consider the case in which the heterotic gauge bundle on $K3$ is
reduced to 24 point-like instantons
\cite{Aspinwall:1997ye,Witten:1995gx}. A point-like instanton is a
limit of a smooth bundle with instanton number 1, in which all the
curvature is concentrated into a single point.\footnote{A point-like instantons can be thought as an NS5 brane wrapping the four-dimensional space-time times the torus $E_H$.}
 If the bundle is given only by point-like instantons, we need 24 of
 them in order to satisfy the Bianchi identity \eqref{Bianchi}. The
 holonomy around the location of a point-like instanton is
 trivial. Hence, the full $E_8\times E_8$ gauge group is unbroken. 
The classical hypermultiplet moduli space is given by the
44-dimensional space 
\begin{equation}
 \Gr_{4,20} \ltimes \mbox{Sym}^{24}(S_H)\ , 
\end{equation}
where the first factor is the $K3$ moduli space, while the second factor
denotes the location of the 24 instantons on $S_H$. Since the
shape of $S_H$ depends on the moduli of the first factor, the product
is a warped product \cite{Aspinwall:2005qw}.
On the vector multiplet side each point-like instanton introduces one
new $d=4$ vector multiplet \cite{Witten:1995gx} so that on the
16-dimensional Coulomb
branch of $E_8\times E_8$ we have altogether $43$ vector
multiplets which
includes  the  three vector multiplets of the heterotic
compactification on $T^2$.

The dual type IIA Calabi-Yau manifold
is a singular elliptic fibration over the Hirzebruch surface
$\mathbb{F}_n$ \cite{Morrison:1996na,Morrison:1996pp}.
$\mathbb{F}_n$ is a $\mathbb{P}^1$ fibration over $\mathbb{P}^1$. It is
described by the homogeneous coordinates $(z,w,u,v)$ identified under the transformations
\begin{equation}
 \mu\, :  \,\,(z,w,u,v) \to (z,w,\mu u,\mu v) \qquad
 \lambda\, : \,\, (z,w,u,v) \to (\lambda z,\lambda w,\lambda^n u, v)\:.
\end{equation}
An elliptic fibration over this space can be described by the Weierstrass
model
\begin{equation}\label{WeieEq}
 y^2 = x^3 + f(z,w,u,v)\,x\,\zeta^4 + g(z,w,u,v) \, \zeta^6 \:.
\end{equation}
For constant $f$ and $g$, this equation describes a torus embedded into the weighted projective space $\mathbb{P}_{1,2,3}^2$. When $f$ and $g$ are functions of the base coordinates, equation \eqref{WeieEq} describes a torus fibration. The elliptic fibre degenerates over points where the discriminant $\Delta = 4 f^3+ 27g^2$ vanishes. If the degeneration is sufficiently hard, this produces a singularity on the Calabi-Yau.
The elliptic fibration over $\mathbb{F}_n$ is also a $K3$ fibration over the $\mathbb{P}^1_{z,w}$ base spanned by the $(z,w)$-coordinates.

In order to make contact with the point-like instanton case, we need
to constrain the complex structure of the Calabi-Yau. In fact, the
corresponding bundle must leave the full $E_8\times E_8$ gauge group
unbroken. On the type IIA side, this is realized by a Calabi-Yau with
two $E_8$ singularities. The corresponding Weierstrass model takes the
form \cite{Morrison:1996na,Morrison:1996pp}
\begin{equation}
 y^2 = x^3 + f_8(z,w) v^4u^4 x + g_{12-n}(z,w) v^5u^7 + g_{12}(z,w) v^6u^6 + g_{12+n}(z,w) v^7u^5 \:,
\end{equation}
with discriminant
\begin{eqnarray}
 \Delta &=& v^{10}u^{10}\left( 4f_8^3 v^2u^2 + 27(g_{12+n} v^2 + g_{12}v\,u + g_{12-n} u^2)^2 \right)\:.
\end{eqnarray}
This space has one $E_8$ singularity at $u=0$ and one $E_8$ singularity at $v=0$. The rest of the discriminant locus intersects these singularities in double points: It intersects $\{v=0\}$ at $12-n$ points (zeroes of $g_{12-n}$), while it intersects $\{u=0\}$ at $12+n$ points (zeroes of $g_{12+n}$).
At these points the singularity becomes worse than an $E_8$
singularity.  
The locations of such singularities on the $\mathbb{P}^1_{z,w}$ base of the $K3$ fibration are controlled by the complex structure of $X$ (related to the coefficients of $g_{12-n}$ and $g_{12+n}$). They correspond to the positions of the heterotic point-like instantons on the base $B_H$. Their positions on the elliptic fibre of $S_H$ are dual to $C_3$ deformations along three cycles of $X_1$ and $X_2$, as explained in Section~\ref{HIIA}.
Looking at the $K3$ fibration structure, the $K3$ fibre has two $E_8$ singularities over each point of the $\mathbb{P}^1_{z,w}$ base.

We can go to the Coulomb branch by resolving the singularities.
Each $E_8$ singularity is resolved by blowing up eight $\mathbb{P}^1$s along the elliptic fibre. They sit at $u=0$ (or $v=0$). Fibering them along the $\mathbb{P}^1_{z,w}$ base of the $K3$ fibration, we get eight exceptional divisors $D_{i,\sigma}^{\tau_{E_8}}$ ($i=1,...,8$ runs over the eight $E_8$ two-cycles and $\sigma=1,2$ runs over the two $E_8$s). On top of the 24 zeros of $g_{12-n}$ and $g_{12+n}$, the singularity is worse than an $E_8$ one. The resolution is made by blowing up the points $\{v=0,\,\,g_{12-n}=0\}$ and $\{u=0,\,\,g_{12+n}=0\}$ of the $\mathbb{F}_n$ base. The corresponding exceptional divisors $D_a^{\WW}$ ($a=1,...,24$) are elliptic fibrations over the blown-up two-cycles in $\mathbb{F}_n$.
Before resolving, the only large divisors were the $K3$ fibre $D^\sigma$, the section of the elliptic fibration $D^{\tau_u}$ and the elliptic fibration $D^{\tau_t}$ of the base of the $K3$ fibration. 
After resolving we have in addition $8+8+24=40$ exceptional divisors. The smooth Calabi-Yau $X$ has then $h^{2,1}(X)=h^{1,1}(X)=43$,
i.e.~Euler characteristic $\chi(X)=0$.

$X$ is self-mirror \cite{Aspinwall:2005qw} and then it trivially respects the conjecture  that the mirror of a $K3$ fibration with heterotic dual is also a $K3$ fibration. 
Under mirror symmetry, the complex structure moduli related to the point-like instanton positions $\omega_a$ are mapped to the coefficients of the K\"ahler form along the (Poincar\'e dual of the) exceptional divisors $D_a^\omega$ located at those points.  
On the other hand, the complex structure moduli $\tau^i$ related to the complex structure of the heterotic $K3$ are mapped under mirror symmetry to the coefficients along $D^{\tau_t}$, $D^{\tau_u}$ and $D^{\tau_{E_8}}_{k,\sigma}$ ($k=1,...,8$ and $\sigma=1,2$).
This allows us to give the leading behavior of $g(\tau,\WW)$ anticipated in \eqref{gtau}, using the intersection numbers of the divisors $D^\tau$ and $D^\WW$. 
The elliptic fibration structure of $D^\WW$ implies that their triple self-intersection is zero. Moreover, as they are blown up at different points on the $\mathbb{F}_n$ base, they do not intersect each other.  Their double self-intersection is an elliptic fibration over points of the blown-up $\mathbb{F}_n$ base. These points in general miss the $D^\tau$ divisor, except for $D^{\tau_u}$ that is the class $\mathbb{F}_n$. Hence, the only non-zero triple intersections with two $D^\WW$ divisors are $D^{\tau_u}\cdot D_a^\WW \cdot D_a^\WW$.
The $D^\tau$ divisors are fibrations of the $K3$ two-cycles over the $\mathbb{P}^1_{z,w}$ base. If we intersect two of them, we obtain a curve that is the fibration of the intersection points on the $K3$ fibre over the $\mathbb{P}^1_{z,w}$ base. These points in general miss the locations of the $D^\WW$ divisors. This does not happen for the exceptional $E_8$ divisors $D^{\tau_{E_8}}_{k,\sigma}$, as they are located at $u=0$ and $v=0$, where the $D^\WW$ divisors sit. In particular, the $12-n$ $D^\WW$ divisors located at $v=0$ intersect the $D^{\tau_{E_8}}_{k,\sigma}$ divisors with $\sigma=1$, while the $12+n$ ones located at $u=0$ intersect the $D^{\tau_{E_8}}_{k,\sigma}$ divisors with $\sigma=2$.

We are now able to give $g(\tau,\WW)$  more explicitly:
\begin{equation}\label{gpi}
 g(\tau,\WW) = \rmi\, \tau^u \sum_{a=1}^{24} \WW^a\WW^a 
		  +\sum_{a=1}^{12-n}  \rmi\WW^a\sum_{k,\ell}\eta^{(1)}_{k\ell}\tau_{E_8}^{k,1}\tau_{E_8}^{\ell,1}
		  + \sum_{a=13-n}^{24}  \rmi\WW^a\sum_{k,\ell}\eta^{(2)}_{k\ell}\tau_{E_8}^{k,2}\tau_{E_8}^{\ell,2}
 + ... \ .
\end{equation}
$\eta^{(\sigma)}_{k\ell}$ is the intersection matrix of the $E_8$ two-cycles of the $K3$ fibre and $\sigma=1,2$ refers to the two $E_8$ singularities. 
We will independently confirm \eqref{gpi} in the next section.

\subsection{Hypermultiplet moduli space from three-dimensional vector multiplets}\label{3dhypermSec}

The hypermultiplet moduli space is not affected by
compactification  
on a further circle 
and in the resulting three-dimensional theory 
we have an additional duality at our disposal.
It descends from a duality in seven space-time dimensions which relates
M-theory compactified on
$K3$ to the heterotic string compactified on $T^3$
\cite{Witten:1995ex}.
Compactifying this duality on another $K3$ to three dimensions one
obtains a relation between two different heterotic theories and two different type IIA theories: 
\begin{equation}\label{HetMthHetdual}
\begin{array}{ccccc}
\mbox{Het}/ (S_H\times \tilde{T}^2 \times \tilde{S}^1) 
 & \leftrightarrow &
\mbox{M-th}/(S_H\times \tilde{S}_H)
   & \leftrightarrow &
\mbox{Het}/( S^1\times T^2 \times \tilde{S}_H) \\[1ex]
\updownarrow &&&& \updownarrow \\[1ex]
\mbox{IIA}/(X  \times \tilde{S}^1) &&&& \mbox{IIA}/( S^1\times\tilde{X}) 
\end{array}\end{equation}
where we take the $K3$ surfaces $S_H$ and $\tilde{S}_H$ 
to be elliptically fibred. 
Starting from M-theory in the center,
the duality on the left first uses the above mentioned 
seven-dimensional duality between M-theory and the heterotic string 
for $\tilde{S}_H$ and then in a second step the 
heterotic--type IIA duality in four space-time dimensions.
On the right the seven-dimensional duality is used 
for ${S}_H$ and as a consequence in the type IIA background 
the mirror Calabi-Yau $\tilde{X}$ appears. 
In all cases the moduli space of the effective three-dimensional theories 
is the product of two quaternionic-K\"ahler spaces. 
Decompactifying on a circle one of the two spaces is projected 
to a special K\"ahler manifold 
and we recover the product of the four-dimensional hypermultiplet 
and vector multiplet moduli spaces \eqref{N=2product}.
Going through the correspondence in detail reveals that the hypermultiplet
moduli space of the heterotic theory on the left is mapped to the
vector multiplet moduli space of the heterotic theory 
on the right and viceversa via the c-map 
\cite{Aspinwall:2005qw,Halmagyi:2007wi}. 
Including the type IIA theories yields the following relations between the respective moduli  spaces: 
\begin{equation}\begin{aligned}\label{MHetMthHetdual}
 {\M}_{\rm  v}(X)\ =\  {\M}_{\rm  v}(S_H\times \tilde{T}^2)
&\quad\leftrightarrow\quad
{\M}_{\rm  h}(\tilde{S}_H \times  T^2)\  =\  {\M}_{\rm  h}(\tilde X)\ ,\\
{\M}_{\rm  h}(X)\ =\ {\M}_{\rm  h}(S_H\times \tilde{T}^2)
&\quad\leftrightarrow\quad
 {\M}_{\rm  v}(\tilde{S}_H \times  T^2)\  =\  {\M}_{\rm  v}(\tilde X)\ .
\end{aligned}
\end{equation}
 Hence, one can derive the hypermultiplet 
moduli space of the heterotic theory in the background  $S_H\times
\tilde{T}^2$ by considering the vector multiplet moduli space of
the  heterotic string in the background $T^2 \times \tilde{S}_H$ and compactifying on a circle $S^1$. This is the `heterotic version' of
deriving the IIA hypermultiplet moduli space on
$X$ by considering the vector multiplet moduli space of type IIA on
$\tilde{X}$, compactifying 
on an circle and adding the quantum corrections. 
This also implies that if the dual of the left heterotic theory is
 type IIA compactified on the CY  $X$ than the dual of the right heterotic theory is  type IIA compactified on the mirror $\tilde{X}$ 
(see \cite{Aspinwall:2005qw}).
Therefore the  duality \eqref {HetMthHetdual} supplies another argument in support of the
conjecture which we stated in section \ref{hetIIAdualresults}:
In order that the dualities on both sides hold $X$ and $\tilde{X}$ 
have to be simultaneously  $K3$ fibrations.

{} From the duality \eqref {HetMthHetdual}
we can also obtain the prepotential as an
expansion in the smoothing parameter $\sigma$ (or the fibre size
$s$). As we reviewed in section~\ref{HIIA} 
the vector multiplet moduli space of the heterotic string 
on $T^2\times \tilde{S}_H$ is given as an expansion in the dilaton $S$
(see \eqref{Fvector}). 
Under the duality \eqref{HetMthHetdual},
this modulus is mapped to the fibre size $s$ of $S_H$ and then to the
smoothing parameter $\sigma$ of type IIA on $X$. Hence, the prepotential of the
vector multiplet moduli space of 
the heterotic theory on $T^2\times \tilde{S}_H$ gives the 
{\it prepotential of the type IIA hypermultiplet moduli space at
  leading order in the parameter $\sigma$ which controls the stable degeneration}.  %

\subsubsection*{Point-like instantons}

Let us apply these considerations to the situation
where the bundle is given by 24 point-like instantons on the smooth $K3$ $S_H$. 
In this case, the dual heterotic background is the same, i.e.~24
point-like instantons on a smooth $K3$ $\tilde{S}_H$. Hence, both are
dual to type IIA on a CY three-fold
which is self-mirror and has $h^{2,1}=h^{1,1}=43$.

The prepotential on the heterotic side is given by the expression \eqref{Hvector}. When the gauge bundle consists of 24 pointlike instantons, we can be more explicit about the functions $f(T)$ and $g(T,W)$. First of all, following the duality \eqref{HetMthHetdual}, one sees that while $S$ corresponds to the complexified volume $s$ of the $S_H$ fibre, the $T$ moduli (that on the vector multiplet side are the complex structure $T_{18}$ of $T^2$,  the complexified volume $T_{17}$ of $T^2$ and the 16 Wilson lines) correspond to the complex structure of $S_H$ that preserves the elliptic fibration, and $W$ to the positions of the pointlike instantons.

Ref.~\cite{Louis:1996mt} determined certain properties of 
$f(T)$ and $g(T,W)$. For example, $f(T)$ is
shown to be a cubic polynomial in the `large' $T$ limit. This
corresponds to keeping the complex structure of $S_H$ generic in order
to avoid shrinking two-cycles, that may introduce $\alpha'$
corrections. In the large $S$ limit
$g(T,W)$ also is a cubic polynomial, that in our case reduces to:
\begin{equation}\label{gTWhethet}
  g(T,W) = -\tfrac12 U \sum_i W_i^2 + \sum_i  \sum_{m=1}^{16} \gamma_m^{i}W_i T_m^2\:.
\end{equation}
When Re$T_{17}>$Re$T_{18}$ holds we have to identify $U=T_{18}$ 
while for  Re$T_{18}>$Re$T_{17}$ we have $U=T_{17}$.
By translating this result in terms of the complex structure of $S_H$ and the positions of the pointlike instantons, we obtain the same expression as in \eqref{gpi}.
In particular, the moduli $U$ and $T_m$ can be identified with
specific complex structure deformations of $S_H$ 
(see \cite{LopesCardoso:1996hq} for details).\footnote{Heterotic
  string 
theory on $K3\times T^2$ with point-like instantons is dual to type
IIB on $K3\times T^2/\mathbb{Z}_2$ orientifold with 16 D7-branes
wrapping $K3$ and 24 D3-branes. The vector multiplet moduli space
prepotential has been worked out in  \cite{Angelantonj:2003zx} at the
classical level. 
Applying the duality map (worked out in detail, for example, in
\cite{Valandro:2008zg}), one recognizes the first term in
\eqref{gTWhethet}, where in type IIB $W_i$ are the positions of the
D3-branes on $T^2$ and $U$ the complex structure of the torus. The
second term should describe a possible coupling between the D3 and the D7
position moduli, arising at quantum level in type IIB.}

\subsubsection*{$SU(2)$ bundle from 3d Seiberg-Witten}

In heterotic string theory compactified on $K3\times T^2$, the hypermultiplet moduli space has a fibration structure, where the base is spanned by the $K3$ moduli and the 
fibre by the bundle moduli. The fibre is known to be an hyper-K\"ahler space (see for example \cite{Aspinwall:2000fd}). 
In this section we 
consider the situation with an $SU(2)$ bundle on the smooth $K3$ $S_H$
and determine the metric on the fibre  at generic values of
the $K3$ moduli. To do so, we freeze the geometric moduli at a
generic point  by taking the volume of $K3$ to infinity.  
The unbroken gauge group (before switching on Wilson lines) is
$E_7\times E_8$. The dual type IIA CY three-fold $X$ 
has $h^{1,1}(X)=18$ and $h^{1,2}(X)=64$ resulting in a $65$-dimensional 
hypermultiplet moduli space \cite{Kachru:1996pc,Candelas:1996su}. Applying 
the duality \eqref{HetMthHetdual} and decompactifying along $S^1$,
one gets a vector multiplet moduli space of (complex) dimension $64$. 
Whenever the $K3$ manifold develops and ADE singularity on one side of the duality, the dual heterotic theory will have a non-abelian gauge bundle with the corresponding structure group.
Since there is an $SU(2)$ bundle on $S_H$, the dual surface $\tilde{S}_H$ has an $A_1$-singularity \cite{Aspinwall:2005qw}. On the other hand, since $S_H$ is smooth, the dual heterotic theory has unbroken $E_8\times E_8$ (perturbative) gauge group (before switching on Wilson lines)
and the Bianchi identity is saturated by 24 point-like instantons. 
Each point-like instanton gives rise to a tensor multiplet in six space-time dimensions, that becomes a vector multiplet after compactifying on $T^2$. When the point-like
instanton sits at a singular point of the $K3$ manifold, new (non-perturbative) vector multiplets arise. In \cite{Aspinwall:1997ye} the
gauge enhancement when $k$ point-like instantons sit on an $A_{1}$-singularity 
is determined to be  $SU(2)^{k-3}$. 
In our situation, the resulting gauge group (excluding the $U(1)$
factor coming from the graviphoton) therefore is
\begin{equation}
 \left(U(1)^3 \times E_8\times E_8\right)\,\times\, SU(2)^{k-3} \,\times\, U(1)^{24} \:.
\end{equation}
The first factor is the perturbative unbroken gauge group; the second
is the enhancement due to 
$k$ instantons on the $A_1$-singularity; the last factor corresponds to
24 vector multiplets associated with the 24 point-like instantons. 
Thus  the dimension of the Coulomb branch 
is $3+8+8+(k-3)+24=40+k$. The expected dimension (i.e. $64$) is
obtained when $k=24$ which means that all the point-like instantons
have to be on top of the $A_1$ singularity. 

If we freeze the vector multiplet moduli dual to the $S_H$ moduli, we are left with the
non-perturbative factor, that is mapped to the bundle moduli space.
Hence, we claim that the bundle moduli space (at $\alpha'\rightarrow 0$) can be reconstructed by the moduli space of the three-dimensional Seiberg-Witten theory 
\cite{Seiberg:1996nz} with gauge group $SU(2)^{21}\times U(1)^{24}$.

There is more: The bundle moduli space of the heterotic theory compactified on $S_H\times \tilde{T}^2$ is described by the deformations of a Riemann surface ${\cal C}_H$ (the spectral curve) together with its Jacobian $J({\cal C}_H)$ \cite{Friedman:1997yq}.
When the bundle has structure group $SU(2)$ with integrated second
Chern class $c_2=24$, the genus of this curve is $g_{{\cal C}_H}=45$ and the curve is a double cover of the $\mathbb{P}^1$ base of $S_H$. 
On the other hand, we have just seen that the same moduli space in the
rigid limit is dual
to the moduli space of a three-dimensional supersymmetric Yang-Mills theory. This space is also described 
by a Riemann surface (the Seiberg-Witten curve) together with its Jacobian. This Riemann surface is a double cover of the complex plane. When the gauge group is $SU(N)$, the genus of the Seiberg-Witten curve is
$g_{SW}=N-1$. 
A match at the level of the curves arises when $g_{SW}=45$, i.e.\ for the gauge group $SU(46)$.
Note that the group $SU(2)^{21}\times U(1)^{24}$ that we found above,
is a subgroup of $SU(46)$ with same rank.
One then would expect points in the vector multiplet moduli space of 
the heterotic theory compactified on $T^2 \times \tilde{S}_H$, where this enhancement is realized. These points should be in a region of the moduli space away from the studied regime. It would be nice to give an interpretation of this enhancement in terms of the dual hypermultiplet moduli space.

\section{Conclusions}

In this paper we considered the duality between heterotic string theory 
compactified on $K3\times T^2$ with the $K3$ being elliptically fibred
and type IIA  compactified on a Calabi-Yau three-fold $X$ with $X$ being both an elliptic and a 
$K3$~fibration. 
We focused on the hypermultiplet sector of the
duality which has not been studied extensively so far.
We analyzed  a region of the moduli space where the metric is given by classical computations on both sides of the duality.
More precisely, we negelected perturbative and non-perturbative quantum
corrections in the string coupling $g_s$ on the type IIA side. 
In this limit 
the hypermultiplet metric is of the Ferrara-Sabharwal form \eqref{qKmetric} and described by a holomorphic prepotential ${\cal G}$. On the heterotic side the hypermultiplet moduli space is exact in string perturbation theory but does receive $\alpha'$ corrections.
They can be negelected for elliptically fibred $K3$'s with both the base of the fibration and the elliptic fibre large.
On the type II side this 
corresponds to taking a specific complex structure modulus (which we denoted
as the smoothing parameter $\sigma$) large. 
In this  region of the moduli space we were able to match the field variables,
the holomorphic prepotentials and thus
the respective effective actions of the two dual theories.
In order to do so we first had to give the hypermultiplet metric of
the $K3$ moduli space in the Ferrara-Sabharwal form and then,
with the resulting 
heterotic prepotential at hand, we were able to explicitly perform the match. 
Furthermore, from the form of the prepotential 
we were led to the conjecture that  type IIA theory compactified on the Calabi-Yau threefold $X$ has a heterotic dual  if not only $X$ but
simultaneously its mirror $\tilde{X}$ is a $K3$ fibration. Mirror symmetry then gives the leading terms of the prepotential ${\cal G}$ in terms of the four-cycle intersection numbers of the mirror $\tilde{X}$. 

Apart from the $K3$ moduli the heterotic prepotential also depends 
on gauge bundle moduli. Using properties of the type IIA theory we identified 
additional terms on the heterotic side, some of which correspond
to the leading $\alpha'$ corrections. 
We also considered in detail the case when the heterotic gauge bundle on $K3$ reduces to 24 point-like instantons. The dual type IIA is compactified on a self-mirror Calabi-Yau manifold with Hodge numbers $h^{1,1}=h^{1,2}=43$.
In this case we have been able to compute the structure of the intersection numbers that enter in the prepotential. This gave us the explicit dependence of  ${\cal G}$ on the bundle moduli.

We confirmed our results by using the duality between  M-theory compactified 
on $K3$ and the heterotic theory compactified on $T^3$. Applying this duality 
twice we  obtained information about the hypermultiplet moduli space of the heterotic theory compactified on $K3\times T^2$ from the vector multiplet moduli space of the heterotic theory also compactified on $K3\times T^2$ --
albeit with a different geometric and gauge background.
Using this correspondence we were also able to derive the hyper-K\"ahler metric on the moduli space of an $SU(2)$ bundle on $K3$.

{}From our analysis it seems feasible to understand in detail
how the heterotic/type IIA duality has to be modified once 
$\alpha'$ and $g_s$ corrections are included. 
We can go away from the `double classical' limit in two ways (see also appendix~\ref{limits}). First of all we can consider subleading  corrections in $g_s$ on the type II side keeping the heterotic $\alpha'$ corrections suppressed.
On the type II side this corresponds to allow $g_s$ to be ${\cal O}(1)$ 
while keeping the smoothing parameter $\sigma$ large.
On the heterotic side this corresponds to having the base volume $\phet$ of the elliptic $K3$  small 
and the fibre volume $s$ large. In this case the hypermultiplet metric is no longer of the Ferrara-Sabharwal form and thus cannot be described by a holomorphic ${\cal G}$. 
On the other hand, since we are suppressing the $\alpha'$ corrections, we obtain the heterotic classical metric at large $K3$ volume.
The second option is to consider subleading  corrections in $\alpha'$ on the heterotic side keeping the $g_s$ corrections on the type II side suppressed. This corresponds to $\phi$ large but taking subleading corrections in $\sigma$ or correspondingly $\phet$ large and $s$ small on the heterotic side. In this case the hypermultiplet metric is of the Ferrara-Sabharwal form and a holomorphic 
${\cal G}$ exists.
Hence, there are regions in the moduli space $\M_{\rm  h}$ where only the classical heterotic metric is corrected by quantum effects, and regions where only the type IIA metric is corrected. This would allow us to understand quantum corrections on one side using the classical results valid on the other side, and viceversa.
In particular, quantum corrections to ${\M}_{\rm  h}$ have been better understood on the type II side of the duality \cite{RoblesLlana:2006ez,RoblesLlana:2006is,Alexandrov:2007ec,RoblesLlana:2007ae,Gaiotto:2008cd,Alexandrov:2008gh,Collinucci:2009nv,Pioline:2009ia}. 
It would be interesting to identify
the heterotic counterpart of these corrections. 
They will give information both on the classical heterotic metric and
on some heterotic quantum corrections. Moreover, considering the
degeneration limit of $X$ on the type IIA side one obtains information
on the classical heterotic metric.
Including the $g_s$ corrections allows to explore regions of the moduli space where the $K3$ base is small, even though the volume is large and then the $\alpha'$ corrections are negligible. 
In this respect, the point-like instanton case that we analyzed in
detail 
in this paper is promising. In fact the dual type IIA Calabi-Yau has $\chi(X)=0$. Since the perturbative $g_s$ corrections are proportional to the Euler characteristic,  the only $g_s$ corrections are the non-perturbative ones. This makes it easier to include the quantum corrections on the type IIA side (in the large $\sigma$ limit) and obtain the full classical heterotic metric.

\vskip 1cm

\vskip 1cm

\subsection*{Acknowledgments}

This work was supported by the German Science Foundation (DFG) within
the Collaborative Research Center (SFB) 676. We have greatly benefited
from conversations and correspondence with S.~Alexandrov, A.~P.~Braun, P.~Candelas, N.~Halmagyi, I.~Melnikov, H.~Ooguri, D.~Persson, B.~Pioline, S.~Sethi and U.~Theis.
We are particularly  indebted to Bernd Siebert for his contribution
and collaboration at an
early stage of this project.

\vskip 1cm

\newpage

\appendix
\noindent
{\bf\Large Appendix}

\section{$K3$ moduli space metric in Ferrara-Sabharwal form}\label{FSf}

In this section we supply the details (i.e.\ the field redefinitions)
which put the metric \eqref{Lhet} into the Ferrara-Sabharwal form 
\eqref{qKmetric}.
For convenience we repeat the action  \eqref{Lhet}
\begin{equation}\begin{aligned}
  \label{Lheta}
{\cal L}\  =\ & -\tfrac18(\partial \rho)^2 + \tfrac1{16}
(\partial \hat{M})^2 -\tfrac14 e^{\rho} \hat{M}^{IJ}\partial
B_I\partial B_J\ ,
\qquad I,J = 0,\ldots, 21\ ,
\end{aligned}\end{equation}
where $e^{-\rho}$ is the $K3$ volume, 
$\hat M$ is a $\SO(3,19)$ matrix parametrizing  $\Gr_{3,19}$
 and $B_I$ are  $22$ $B$-fields.\footnote{In this appendix we do the
computation for all $K3$ moduli but one could easily freeze some of
them to obtain the results for a subspace.} 

As a first step we recall a more explicit representation
of $\hat M$ in terms of scalar fields that was
determined in \cite{Louis:2009dq}. Since $K3$ is a
hyper-K\"ahler manifold there exist three complex structures $J^x, x=1,2,3$.
One can define their deformation $\zeta$ via
\begin{equation}
 \label{defJ}
J^x = e^{-\frac12\rho}\zeta_I^x \omega^I\ ,\quad 
\end{equation}
where $\omega^I$ are the 22 two-forms of $K3$.
The 66  $\zeta_I^x$ are constrained by the six equations
\begin{equation}
 \label{zetacon}
J^x\wedge J^y = 2 \delta^{xy}\, {\rm vol}(K3)\quad\Rightarrow\quad
\eta^{IJ} \zeta_I^x \zeta_J^y = 2\delta^{xy}\ ,
\end{equation}
where the intersection matrix
\begin{equation} \label{etadef}
\eta^{IJ} = \int_{K3} \omega^I\wedge\omega^J
\end{equation}
has signature $(3,19)$. Furthermore the $\SO(3)$ rotation among the $J^x$
subtracts additionally 3  angles so that 
there are $66-6-3=57$ independent parameters which can be viewed as the
coordinates of $\Gr_{3,19}$.
In terms of the $\zeta_I^x$ the matrix $\hat{M}$ takes the form \cite{Louis:2009dq}
\begin{equation}
 \label{Mzeta}
\hat{M}_{IJ} = -\eta_{IJ} + \zeta_I^x \zeta_J^x \ .
\end{equation}

For our purpose it is convenient to choose a specific basis where 
we can easily identify the K\"ahler moduli corresponding
to the base and fibre of the elliptic $K3$. Furthermore we also want 
 to 
choose a parametrization of $\Gr_{3,19}$ which treats the 
K\"ahler moduli as  fibred over the base  $\Gr_{2,18}$
which is spanned by the complex structure deformations preserving the elliptic fibration.
To do so we split the basis $\{\omega^I\}$ 
into two sets of elements which are orthogonal to each other 
with respect to the metric \eqref{etadef}. 
The first set consists of the two elements $\omega^0,\omega^1$ 
corresponding to the $\mathbb{P}^1$-base and the elliptic fibre of
the $K3$. The second set contains the others 19 elements $\omega^A$.
In this basis  the intersection matrix \eqref{etadef} splits into two blocks:
\begin{equation}
 \label{etadef2bl}
\eta^{\alpha\beta} =\left(\begin{array}{cc} 0&1\\1&0\\
   \end{array}\right) \ ,\qquad \quad
\eta^{AB} = \int_{K3} \omega^A\wedge\omega^B\ , \qquad A=2,...,21\ .
\end{equation}

Furthermore let us fix the $SO(3)$ redundancy by choosing 
$\Omega = J^1+iJ^2$ for 
the holomorphic two-form $\Omega$ and 
$J^3$ for the real K\"ahler form.
Now the constraints \eqref{zetacon} can be explicitly 
solved by introducing the following parametrization
\begin{equation}\label{zetatrans}
 \begin{array}{lclcl}
  \zeta_0 = -\eta^{AB} \zeta_A \ttt_B \ ,&& \zeta_1 = 0  \ ,&& \zeta_A = \frac{X_A}{\sqrt{\eta^{BC}X_B\bar X_C}} \ , \\ \\
  \zeta^3_0 = e^{-R/2} -  \tfrac{1}{2} e^{R/2} v^Av_A \ , && \zeta^3_1 = e^{R/2} \ ,  && \zeta^3_A = e^{R/2} \ttt_A \ , 
  \end{array}
\end{equation}
where $\eta^{BC} X_BX_C = 0$ and we defined 
$\zeta_I \equiv \frac12 (\zeta^1_I + i \zeta^2_I)$.
Inserted into  $\hat{M}$ of \eqref{Mzeta} we arrive at
\begin{equation}\label{Mpar}
 \hat{M}_{IJ} = \left(\begin{array}{ccc}
 e^{-R}+\frac{\ttt^4}{4}e^R+\ttt_A M^{AB} \ttt_B &  -\frac{\ttt^2}{2}e^R  &  - \ttt^C M_{CB} - \frac{\ttt^2}{2}e^R \ttt_B \\
 -\frac{\ttt^2}{2}e^R  & e^R  &  e^R \ttt_B \\
 - M_{AD}\ttt^D - \frac{\ttt^2}{2}e^R \ttt_A  &  e^R \ttt_A  &  M_{AB} + e^R \ttt_A \ttt_B
 \end{array} \right) 
\end{equation}
with $I=\{\alpha,A\}$ and the indices $A,B$ raised by $\eta^{AB}$.
The matrix $M_{AB}$ takes the form
\begin{equation}
 \label{MX}
M_{AB} = -\eta_{AB} + 2\, \frac{X_A\bar X_B + \bar X_A
  X_B}{\eta^{CD}X_C\bar X_D}\ .
\end{equation}

Note that the field redefinitions \eqref{zetatrans} express the 66 $\zeta^x_I$
in terms of 20 complex $X^A$, 20 real $\ttt_A$ plus one real $R$.
However, the constraint  $\eta^{BC} X_BX_C = 0$ removes two degrees of
freedom and we see from \eqref{MX} that the overall complex scale of
the $X^A$ does not appear in $\hat M$. Thus effectively 
the $X_A$ depend on 18 complex structure moduli $t^i, i=1,\ldots,18$ 
which  preserve the elliptic fibration.  In addition 
there are the 22 (real) K\"ahler moduli $\ttt_A,R,\rho$ adding up to 58
$K3$ moduli altogether.

Inserting \eqref{Mpar} into \eqref{Lheta} we obtain
\be\ba\label{HetMetricInterm}
 {\cal L} =& -\tfrac18 (\partial \rho)^2 -\tfrac18 (\partial R)^2 + \tfrac{1}{16} (\partial M)^2 -\tfrac14 e^R M^{AB}\partial \ttt_A \partial \ttt_B 
  -\tfrac14 e^\rho M^{AB}\partial B_A \partial B_B\\
   & -\tfrac{1}{4}e^\rho \left[e^R\left( \partial b - \tfrac{\ttt^2}{2}\partial b_* + \ttt\cdot \partial B \right)^2 + \partial b_* \left(
  (\ttt\cdot M \cdot \ttt) \partial b_* - 2 \ttt\cdot M\cdot \partial B
\right)\right] \\
&-\tfrac{1}{4}e^{\rho-R}(\partial b_*)^2\ ,
\ea\ee
where we used the notation 
$B_1 = b$ and $B_2 = b_*$.
As a next step let us define 
\be\ba
\chi & =\tfrac14 (\rho+R)\ ,\qquad
a= - \tfrac{1}{2}(b + \tfrac12 \ttt^CB_C)\ ,\qquad
s=\nu_F+\rmi b_*\ ,\\
\tilde{c}_A &=\tfrac{1}{\sqrt{2}}(B_A-b_* \ttt_A)\ ,\qquad
c^A =\tfrac{1}{2\sqrt{2}}\, \ttt^A\ ,\\
\ea\ee
where $\nu_F=e^{-\tfrac12(\rho-R)}$ is the volume of the fibre.
These redefinitions put the metric \eqref{HetMetricInterm} into the form:
\be\ba\label{LhetCmap}
 {\cal L} &= -(\partial \chi)^2 
    -\tfrac14{e^{4\chi}}\left(\partial a +\partial c^A \tilde c_A
- c^A \partial\tilde c_A \right)^2 
  -\frac{\partial s \partial \bar{s}}{(s+\bar{s})^2}
+ \tfrac{1}{16} (\partial M)^2  \\
  & -e^{2\chi}\,\frac{M^{AB}}{s+\bar{s}}\left[ \partial\tilde c_A - 
{\cal N}_{AD}\partial c^D\right]
\left[ \partial\tilde c_B 
- \overline{{\cal N}}_{BC}\partial c^C\right]\ ,
\ea\ee
where
\be\label{cM}
{\cal N}_{AD}= - {\rmi}\left[({s-\bar{s}})\eta_{AD} + ({s+\bar{s}})M_{AD}\right]\ .
\ee
This is almost of the form \eqref{qKmetric}. The only subtlety left is the 
fact that the $X^A$ are not all independent (they obey  $X_AX^A=0$)
and therefore they cannot yet be viewed as
homogeneous coordinates of a special K\"ahler manifold.
In fact the parametrization \eqref{LhetCmap}, \eqref{cM} corresponds to
a field basis of the special K\"ahler base where the holomorphic
prepotential $\cG$ does not exist \cite{Ceresole:1995jg}.
This basis is frequently encountered in heterotic compactifications 
\cite{deWit:1995zg,Louis:2001uy} and can be related to the standard
basis as given in \eqref{gdef}, \eqref{Ndef} by a symplectic rotation.

\begin{table}[t]\begin{center}
\begin{tabular}{|ccc|c|}
\hline
IIA& & Het& $\#_{\mathbb{R}}$\\
\hline\hline
$\Phi=e^{-\phi} + \rmi\,\ax$ & $=$ & $\phet=e^{-\chi} +\rmi a$&2\\
$\sigma$& $=$ &  $s=\nu_F+\rmi\, b_*$&2\\
$\tau_i$& $=$ &  $t_i$ &36\\
$\tilde{\xi}_A$& $=$ & $\tilde c_A$&20\\
$\xi^A$& $=$ &  $c^A$ &20\\
\hline
\end{tabular}
\caption{Type II -- heterotic tree-level map}\label{tab:map2}
\end{center}\end{table}

In order to rotate to the standard basis,
let us first note that the constraint $X_AX^A=0$ can be solved by the choice
\begin{equation}\label{Xchoice}
 X_A = \left\{ X^0, X^0\eta^{ij} t_i t_j, \rmi X^0 t_i \right\} \ ,\quad
 i,j=1,...,18\ ,
\end{equation}
where we splitted 
\be
\eta_{AB}=\begin{pmatrix} 0 & 1/2&0\\1/2 & 0 &
   0\\0&0&\eta_{ij}\end{pmatrix}\ .
\ee
Inserted into \eqref{MX} one obtains \cite{Louis:2001uy}
\begin{equation}\label{dMdMhet}
 \tfrac{1}{16}(\partial M)^2\ =\ \frac{\partial}{\partial
   t_i}\frac{\partial}{\partial \bar{t}_j} K(t,\bar t)
\ ,
\ee
with
\be
K= -\log\left(\frac{2X_A\eta^{AB}\bar{X}_B}{|X^0|^2}\right)
    = -\log\left(\eta_{ij}(t^i+\bar{t}^i)(t^j+\bar{t}^j)\right)\ .
\end{equation}
Together with the metric for $s$ from \eqref{LhetCmap} we have
\begin{equation}
K= -\log(s+\bar s) -\log\left(\eta_{ij}(t^i+\bar{t}^i)(t^j+\bar{t}^j)\right)\ ,
\end{equation}
which indeed is the K\"ahler potential of the special K\"ahler
manifold
\be
\M_{\rm SK}=\frac{SU(1,1)}{U(1)}\times\Gr_{2,18}\ ,
\ee
with a  prepotential \eqref{prepMir}, i.e.\ $\cG_{\rm het} = \rmi s \eta_{ij}t^it^j$.
The symplectic rotation
\begin{equation}\label{SymplS}
(Z^s,\cG_s)\ \mapsto\ (\cG_s,-Z^s)
\end{equation}
 precisely maps to the field basis
\eqref{Xchoice} and furthermore transforms the matrix ${\cal N}$
defined in \eqref{Mdef} into the form \eqref{cM} 
\cite{Ceresole:1995jg,deWit:1995zg,Louis:2001uy}.

Thus we finally succeeded in putting the heterotic metric
\eqref{Lheta}
into the Ferrara-Sabharwal form \eqref{qKmetric}.
Therefore we can 
now give  the 
precise (classical) map between type~IIA and heterotic moduli
which is summarized in table~\ref{tab:map2}. 
\footnote{In the last two rows of the table~\ref{tab:map2} we have been schematic: Because of the symplectic transformation \eqref{SymplS}, the pair of fields $(c^s, \tilde c_s)$, which correspond to the index $A=s$, should be swaped.}

\section{Refined classical limits}\label{limits}

Let us first determine the geometrical meaning of some of the moduli
introduced in the previous appendix.
To do so we consider  the K\"ahler-form $J$ evaluated in 
the basis of the previous appendix.  That is we insert
\eqref{zetatrans} into \eqref{defJ} to arrive at
\be
J = J^3=e^{-\frac12\rho}\zeta_I^3 \omega^I=
(e^{-2\chi} -  \tfrac{1}{2}\nu_F v^Av_A)\omega^0+ \nu_F \omega^1
+ \nu_F \ttt_A \omega^A\ .
\end{equation}

In the basis \eqref{etadef2bl}  the elliptic fibre of $S_H$ is dual to 
$\omega^0$, while the $\mathbb{P}^1$-base is dual to  $\omega^1-\omega^0$.
This can be checked by computing the intersection matrix for fibre and
base to be
\begin{equation}
\eta=\left(\begin{array}{rr} 0&1\\1&-2\\
   \end{array}\right) \ .
\end{equation}
We see that the fibre has no self-intersection but it intersects once
with the base. The base itself has self-intersection $-2$ which 
confirms that the base is a $\mathbb{P}^1$ while the fibre
is elliptic.
Now we compute \footnote{
For generic $v^A$, the volume of the base $B_H$ gets a non-zero contribution also from $\int_{K3}\Omega\wedge (\omega^1-\omega^0)$. We are neglecting these contributions here, as they are irrelevant for our conclusions.}
\begin{equation}\ba\label{volumes}
\mbox{vol}(S_H) &=\tfrac12 \int_{K3}J\wedge J= e^{-\rho}\ , \qquad
\mbox{vol}(F_H)=\int_{K3}J\wedge \omega^0= \nu_F = e^{-\tfrac12(\rho-R)}\ ,\qquad\qquad  \\
\qquad\qquad \mbox{vol}(B_H)& =\int_{K3}J\wedge (\omega^1-\omega^0)
= e^{-\tfrac12(\rho+R)} -\tfrac12 e^{-\tfrac12(\rho-R)}( v^Av_A+2)\ .
\ea\end{equation}
We see that $e^{-\rho}$ is the $K3$ volume, $\nu_F$ is
the volume of the fibre while the volume of the base is a more
complicated linear combination.

Let us now discuss various limits we can take.
First of all  the $K3$ volume becomes large in the limit 
$\rho \to -\infty$ and $R$ {\it generic}. 
From \eqref{volumes} we see that in this limit also
the volume of base and fibre are both large which
implies that the $\alpha'$ corrections on the heterotic side
vanish. To see this more explicitly we note that $\alpha'$ corrections
arise from worldsheet instantons wrapping a minimal two-sphere in $K3$
(two-cycles with self-intersection equal to $-2$) with 
the strength of the correction depending on the area of the wrapped
two-sphere. 
For general complex structure of $S_H$, generic two-spheres are
orthogonal to the Piccard lattice and then their area is given by
\begin{equation}
 \mbox{vol}(S^2)=(\mbox{vol}(S_H))^{1/2} \, \left|\int_{S^2}\Omega
 \right|\sim e^{-\rho/2}\ .
\end{equation}
We see that the area is going to infinity when the volume of $S_H$ is large and the complex structure is generic. 
Thus the $\alpha'$ corrections  vanish on the heterotic side
in the limit $\rho \to -\infty$ and $R$ {\it generic}.
From table~\ref{tab:map2} we further conclude that 
in the dual type II theory $\Phi\sim g_s^{-2}$ becomes large in that
limit and therefore   the type II $g_s$ corrections 
vanish correspondingly.
Or in other words in this limit $\alpha'$ corrections on the heterotic side and $g_s$
corrections on the type II side vanish simultaneously.

However, since the $K3$ is elliptically fibred one can also take two more refined limits where either one of
these corrections are kept.
In the limit $\rho$ finite and $R\to -\infty$ the volume of $S_H$ is
finite while the volume of the base $B_H$ is still
infinite. Therefore heterotic $\alpha'$ corrections are kept while 
the $g_s$ corrections continue to vanish. Note that in this limit, the
IIA CY three-fold $X$ goes away from the degenerate limit $X^\sharp$ which
has been discussed in \cite{Aspinwall:1999xs}.

The other possible limit is 
$\rho \to -\infty$, $R\to \infty$, $e^{-\tfrac12(\rho+R)}$
   finite and $v^Av_A=-2$. 
The volume of $S_H$ and $F_H$ is infinite  while the volume
of $B_H$ is finite. This corresponds to negligible heterotic $\alpha'$
corrections but keeping the type II $g_s$-corrections.
Note that in this limit the IIA CY is the degenerate $X^\sharp$.


\end{document}